\documentclass[12pt,a4paper]{article}

\usepackage[margin=1in]{geometry}
\usepackage[authoryear,round]{natbib}
\usepackage[strings]{underscore}
\usepackage{amsfonts,amssymb,amsmath}
\usepackage{graphicx}

\usepackage[margin=1cm]{caption}
\usepackage{multirow}
\usepackage{longtable}

\title{Cryptographically secure multiparty evaluation of system reliability}
\author{Louis~J.~M.~Aslett \\ {\normalsize Department of Statistics, University of Oxford}}
\date{}

\newcommand{\Enc}{\texttt{Enc}}
\newcommand{\Dec}{\texttt{Dec}}

\begin{document}
\maketitle

\begin{abstract}
  The precise design of a system may be considered a trade secret which should be protected, whilst at the same time component manufacturers are sometimes reluctant to release full test data (perhaps only providing mean time to failure data). In this situation it seems impractical to both produce an accurate reliability assessment and satisfy all parties' privacy requirements. However, we present recent developments in cryptography which, when combined with the recently developed survival signature in reliability theory, allows almost total privacy to be maintained in a cryptographically strong manner in precisely this setting. Thus, the system designer does not have to reveal their trade secret design and the component manufacturer can retain component test data in-house.
  
\end{abstract}

\section{Introduction}\label{sec:Introduction}

A standard part of developing any critical system is a thorough assessment of the reliability characteristics.  This involves a careful assessment of the reliability of the components which will comprise the system and then propagation of all the uncertainty in these estimates through to an assessment of the reliability of the whole system.

A common impediment to the above best practice is commercial privacy concerns: the system designer will typically consider the exact design of their system to be a trade secret, while component manufacturers are frequently reluctant to make full test data on their components available, often preferring to give summary statistics such as mean time to failure or other similarly suboptimal measures of reliability.  Achieving a full and frank assessment of overall system reliability while respecting such privacy concerns appears to present a significant challenge.

Sensitive data is often kept private by encrypting it.  Traditional methods of encryption are primarily intended for transmission of sensitive information and do not allow any meaningful computation using the encrypted data without first decrypting it and risking revealing everything.  However, recent developments in the cryptography literature \citep{Gentry:2009aa} have provided methods of encryption which allow computation on the encrypted content and has been shown to hold substantial promise in achieving privacy preserving statistical analyses \citep{Graepel:2012aa,Aslett:2015ac}.

Herein these techniques are used to solve the widespread privacy problem of reliability evaluation in industrial settings where privacy of test data and system design are commercial requirements.  Section \ref{sec:HE} provides a high-level introduction to \emph{homomorphic encryption}, avoiding any technicalities that are not directly relevant.  Section \ref{sec:SurvSig} describes the survival signature and shows its importance in this context as a reliability decomposition which is amenable to encrypted computation.  This leads to the privacy preserving protocol described in Section \ref{sec:PrivacyPreservingProtocol}, which enables evaluation of the full survival curve of a system without disclosure of component test data or system design.  There is a practical example discussed in Section \ref{sec:ExperimentalResults} to provide practitioners with a clear demonstration of the practicality of the method.  The paper concludes with a short discussion in Section \ref{sec:Conclusions}.

\section{Homomorphic encryption}\label{sec:HE}

In this section a high-level introduction to homomorphic encryption schemes is provided avoiding any unnecessary technicalities.  Later it will be shown that the form of the survival signature is very special from the perspective of performing homomorphic computations in reliability problems, providing the motivation for the privacy preserving protocol presented in Section \ref{sec:PrivacyPreservingProtocol}.

\subsection{Standard cryptography}

All encryption schemes are either so-called `public' or `secret' key in nature.  The protocol to be presented in the sequel requires a public key scheme.  A \emph{public key} encryption scheme has two algorithms $\Enc(\cdot, \cdot)$ and $\Dec(\cdot, \cdot)$, which perform encryption and decryption respectively, together with two keys: the public key, $k_p$, can be distributed widely and anyone may use it to encrypt a message; the secret key, $k_s$, is required to decrypt any message encrypted using $k_p$ and so kept private.  The principal identity is:
\[ \Dec(k_s, \Enc(k_p, m)) = m \quad\forall\,m \]
The data to be encrypted, $m$, is referred to as the \emph{message} and after transformation by $\Enc(\cdot, \cdot)$ is referred to as the \emph{ciphertext}.

Encryption is a common technique for ensuring the privacy of data, but typically once one wishes to perform an analysis it is necessary to first decrypt and risk exposing the data.  In \citeyear{Rivest:1978aa} \citeauthor{Rivest:1978aa} proposed that encryption schemes capable of theoretically arbitrary computation on encrypted data may be possible, but it was not until \citeyear{Gentry:2009aa} that \citet{Gentry:2009aa} provided the first such scheme.  An explosion of advances in cryptography have ensued, each enabling a limited amount of computation to be performed directly on the encrypted content, rendering the correct result upon decryption.

\subsection{Fully homomorphic schemes}

Encryption schemes are said to be \emph{fully homomorphic} if they also possesses two operations, $\oplus$ and $\otimes$, that can be applied a theoretically arbitrary number of times and satisfy:
\begin{align*}
  \Dec(k_s, \Enc(k_p, m_1) \oplus \Enc(k_p, m_2)) &= m_1 + m_2 \\
  \Dec(k_s, \Enc(k_p, m_1) \otimes \Enc(k_p, m_2)) &= m_1 \times m_2
\end{align*}
for all $m_1, m_2$.  In other words, a homomorphic encryption scheme allows computation directly on ciphertexts, which will correctly decrypt the result as if it had been applied to the original messages.

However there are many constraints in practical implementation, reviewed in \cite{Aslett:2015ab}.  For the purposes of this work they may be synopsised as:
\begin{enumerate}
  \item Typically $m$ can only naturally represent binary ($m \in \mbox{GF}(2)$) or limited integer ranges ($m \in \mathbb{Z}/n\mathbb{Z}$).
  \item Data size after encryption is substantially inflated, by orders of magnitude.
  \item Computational costs for $\oplus, \otimes$ are orders of magnitude higher than standard $+, \times$.
  \item Comparisons ($=, <, >$) and division are not possible under current schemes.
  \item Implementation of current schemes necessitate a hugely computationally expensive `bootstrap' operation (not related to statistical bootstrapping) which must be applied frequently between $\otimes$ operations to control the noise in the ciphertext.  Indeed, bootstrap operations are so expensive that it is typical to avoid them completely by constraining the algorithm to a small number of successive multiplications.
\end{enumerate}

In essence then, homomorphic schemes are capable of evaluating multivariate polynomials of the ciphertexts.  In other words, if $f(m_1, \dots, m_n)$ is a multivariate polynomial in $m_1, \dots, m_n$ then a homomorphic scheme ensures \[ \Dec\Big(k_s, g\big(\Enc(k_p, m_1), \dots, \Enc(k_p, m_n)\big)\Big) = f(m_1, \dots, m_n) \] where $g(\cdot)$ is the function $f(\cdot)$ with $+$ replaced by $\oplus$ and $\times$ replaced by $\otimes$.  Hence, any algorithm to be computed encrypted must be expressible as a polynomial function.

Accordingly, constraint 5 above implies that the degree of the polynomial to be evaluated is a crucial quantity: if the degree is kept small then bootstrap might be avoidable and the algorithm will be fast.  Herein this quantity is referred to as the \textit{multiplicative depth} of the algorithm.  In particular, note that the total number of multiplications is not the issue, rather the number of successive multiplications.

The final point to note is that all homomorphic encryption schemes involve fixing some parameters before generating the public and secret keys.  These parameters influence how secure the scheme is and often have other side effects such as controlling the multiplicative depth which is possible.  Often there are theoretical analyses of homomorphic schemes to enable specification of a desired security and multiplicative depth to then work backwards to an appropriate parameter choice.

Despite such restrictions, fully homomorphic encryption is theoretically exciting because for binary messages $+$ and $\times$ correspond to logical OR and XOR respectively.  These two logical operations are sufficient for construction of any logical circuit and therefore sufficient for theoretically arbitrary computation.

\subsubsection*{Practical usage of homomorphic schemes} \label{sssec:PracticalHE}

It is currently impractical to encrypt all data as binary and compute the corresponding logical circuits because the computational burden is too great \citep[see][for a fuller discussion]{Aslett:2015ab}.  Herein, an integer encoding of real values is used of a nature similar to \citep{Aslett:2015ac}.

Let $y \in \mathbb{R}$, then an encoding $m \in \mathbb{Z}$ of $y$ is required which can be encrypted.  Let $\mu_\kappa: \mathbb{R} \to \mathbb{Z}$ be the encoding function used to represent real valued data as a message with $m = \mu_\kappa(y) \triangleq \lfloor 10^\kappa y \rceil$, where $\lfloor \cdot \rceil$ denotes rounding to the nearest integer.  Then this transformation, parameterised by $\kappa$, preserves $\kappa$ decimal places of the real value $y$ with representation in integer form.  $y$ can be approximately recovered by $\mu_\kappa^{-1}(\mu_\kappa(y)) = 10^{-\kappa} \mu_\kappa(y) \approx y$

This encoding works well for addition and multiplication, since
\begin{align}
  \mu_\kappa^{-1}\big(\mu_\kappa(y_1)+\mu_\kappa(y_2)\big) &\approx y_1 + y_2 \nonumber\\
  \mu_{2\kappa}^{-1}\big(\mu_\kappa(y_1)\mu_\kappa(y_2)\big) &\approx y_1 y_2 \label{eq:IntEncPrec}
\end{align}
Note the precision adjustment, $2\kappa$, required in \eqref{eq:IntEncPrec}.  However, care is clearly needed since encodings with differing precision parameters cannot be summed without adjustment.

\bigskip
The above discussion makes clear that standard statistical methods often cannot be applied unmodified on encrypted content.  However, as will be shown below the form of the survival signature is particularly well suited to encrypted reliability analysis.

For further details, the reader is directed to fuller reviews of homomorphic encryption \citep{Gentry:2010aa,Aslett:2015ab}.


\section{Encrypted computability and survival signatures} \label{sec:SurvSig}

The survival signature was introduced by \citet{Coolen:2012aa} as a generalisation of the system signature \citep{Samaniego:1985aa, Kochar:1999aa}, enabling multiple different types of component.  This relaxed the restrictive assumption of i.i.d components found in the system signature, whilst retaining the separation of component lifetime and the effect of system structure on reliability.  This separation of components and structure is the first critical property of the survival signature which enables the privacy preserving protocol to be implemented in the sequel.

The structure function \citep{Birnbaum:1961aa} of an $M$ component system is the function $\phi : \{0,1\}^M \to \{0,1\}$ which maps the binary operational state of the $M$ components to the operational state of the whole system.  Letting $\mathbf{x} = (x_1, \dots, x_M) \in \{0,1\}^M$ be the vector of component states, $\phi(\mathbf{x})$ is then the non-decreasing structure function representing the system state.  In particular, note that for coherent systems $\phi(\mathbf{x})$ is always a multivariate polynomial in $\mathbf{x}$, so that given $\mathbf{x}$ encrypted the operational state of the system can be computed without decrypting the individual component states.

Indeed, it is valid to replace the binary operational state vector $\mathbf{x}$ with probabilities of correct function at some mission time of interest and the structure function will then render the probability of correct system operation.  However, note that in this instance, if the probabilities are encoded as in \S\ref{sssec:PracticalHE}, then evaluation becomes awkward due to the varying precisions $\kappa$.  Much more problematically, the form of the polynomial to be evaluated is itself an encoding of the system design, so that revealing the algorithm to evaluate the reliability is equivalent to revealing the design, rendering it unusable for the problem considered here.

Let the system in question have $K \ge 1$ different component types, with $M_k$ components of type $k$, so that $\sum_{k=1}^K M_k = M$.  It then follows \citep{Coolen:2012aa} that the probability the system works given that $\mathbf{l}=(l_1, \dots, l_K)$ of each of the $\mathbf{M}=(M_1, \dots, M_K)$ components is working is:
\begin{equation}
  \label{eq:SurvSig}
  \Phi(\mathbf{l}) = \left[ {M_k \choose l_k}^{-1} \right] \sum_{\mathbf{x} \in S_{\mathbf{l}}} \phi(\mathbf{x})
\end{equation}
where $S_{\mathbf{l}}$ is the set of all component states $\mathbf{x}$ with exactly $\mathbf{l}=(l_1, \dots, l_K)$ components of each type working, $S_{l_1, \dots, l_K} = \{ \mathbf{x} : \sum_{i=1}^{M_k} x_i^k = l_k \quad\forall\,k \}$.  Equation \eqref{eq:SurvSig} is called the \emph{survival signature}.

If $C_t^k \in \{ 0, 1, \dots, M_k \}$ is the random variable denoting the number of components of type $k$ which are still operational at time $t$ in a future system with survival signature $\Phi(\cdot)$, then the survival signature allows decomposition of the survival function of the overall system lifetime $T_S$ as:
\begin{align}
  \mathbb{P}(T_S > t) &= \sum_{l_1=0}^{M_1} \cdots \sum_{l_K=0}^{M_K} \Phi(l_1,\ldots,l_K) \mathbb{P}\left(\bigcap_{k=1}^K \{C^k_t = l_k\}\right) \nonumber\\
  &= \sum_{l_1=0}^{M_1} \cdots \sum_{l_K=0}^{M_K} \Phi(l_1,\ldots,l_K) \prod_{k=1}^K \mathbb{P}\left(C^k_t = l_k\right) \qquad \mbox{if types are independent} \label{eq:SurvSigLife}
\end{align}

Notably, \eqref{eq:SurvSigLife} is also a polynomial in $\Phi(\cdot)$ and the component probabilities.  Moreover, it is a \emph{homogeneous} polynomial.  Therefore, under the scaling of \S\ref{sssec:PracticalHE}, every term of the polynomial in \eqref{eq:SurvSigLife} has precision $(K+1) \kappa$, so that the terms can be summed without any rescaling required.  This makes the survival signature particularly well suited to computation under a homomorphic encryption scheme.  Further still, the polynomial form of \eqref{eq:SurvSigLife} is identical for all systems containing $K$ types of component so that it reveals little about the design of the system as long as the values of $\Phi(\mathbf{l}) \ \forall\,\mathbf{l}$ are not revealed.

\bigskip
A recent detailed survey of the survival signature with a more expansive introduction can be found in \citep{Coolen:2016aa}.

\section{The privacy preserving protocol}\label{sec:PrivacyPreservingProtocol}

The exposition hereinbefore highlighted the crucial aspects of homomorphic encryption and the survival signature which now enable a novel approach to preserving the privacy of component manufacturers and system designers when evaluating the reliability of a new system.

\subsection{Privacy framework}

The form of privacy protocol now presented falls under the so-called `honest but curious' model of security without collusion.  That is, it is assumed that all the parties participating in the protocol are honest in following the specification, honest in the data they input and do not collude with one another (rather are in competition).  However, it is assumed that all parties are curious to determine the secret information belonging to the other parties and will seek to learn it by any means possible outside the protocol.

An interesting open question is to extend this to the case of malicious parties, where there are no guarantees that the protocol has been followed and so trust in the final answer must be dynamically established.

\subsection{Setup}

Let the system designer who wishes to evaluate the reliability of their design be called $\delta$.  The design involves $K$ different types of component, made by manufacturers called $\chi_1, \dots, \chi_K$.

There are certain initial setup actions which must be performed:
\begin{enumerate}
  \item $\delta$ first creates public and secret keys $(k_p, k_s)$ for a homomorphic encryption scheme of their choice which supports integer encryption.  The scheme parameters used to generate the keys must support multiplicative depth of at least $K$.
  \item $\delta$ selects a grid of time points at which they wish to evaluate the survival curve of the system, $\mathbf{t} = \{ t_1, \dots, t_T \}$.  
  \item $\delta$ next creates an encrypted lookup table, $\Xi$, for the system design.  Part will be encrypted and part unencrypted:
    \begin{enumerate}
      \item Form a table, $\Xi^{(l)}$, comprising of columns $l_1, \dots, l_K$ with rows enumerating all possible combinations of values that can be taken (i.e.\ $\prod_{k=1}^K (M_K+1)$ rows).
        \[ \Xi^{(l)} = \left( \begin{array}{ccc}
          0 & \cdots & 0 \\
          0 & \cdots & 1 \\
          \ & \vdots & \\
          l_1 & \cdots & l_K \\
          \ & \vdots & \\
          m_1 & \cdots & m_K
        \end{array} \right) \]
      \item Form another table, $\Xi^{(\Phi)}$, containing $T$ identical columns, where all entries on row $i$ are $\Phi\big(\Xi_{i,\cdot}^{(l)}\big)$.  That is, each row contains the survival signature for each corresponding input row in $\Xi^{(l)}$.  Note this requires only knowledge of the system design.
        \[ \Xi^{(\Phi)} = \left( \begin{array}{ccc}
          \Phi(0, \dots, 0) & \cdots & \Phi(0, \dots, 0) \\
          \Phi(0, \dots, 1) & \cdots & \Phi(0, \dots, 1) \\
          \vdots & & \vdots \\
          \Phi(l_1, \dots, l_K) & \cdots & \Phi(l_1, \dots, l_K) \\
          \vdots & & \vdots \\
          \Phi(m_1, \dots, m_K) & \cdots & \Phi(m_1, \dots, m_K)
        \end{array} \right) \]
        Although all columns are identical at the start of the protocol, each column will differ as the protocol proceeds once manufacturers interact test data for the $T$ different time points.  That is, each column will be the reliability evaluation for a different level of function at each time $\{ t_1, \dots, t_T \}$.
      \item $\delta$ then decides on an acceptable precision level, $\kappa$, and mutates $\Xi^{(\Phi)}$, \[ \Phi(\Xi_{i,\cdot}^{(l)}) \to \Enc\Big(k_p, \mu_\kappa\Big(\Phi\big(\Xi_{i,\cdot}^{(l)}\big)\Big)\Big), \] so that the lookup columns $l_1, \dots, l_K$ in $\Xi^{(l)}$ are unencrypted and the survival signature values in $\Xi^{(\Phi)}$ are encrypted.  Together these form an encrypted representation of the system structure that reveals the number of components, but keeps the design secret.
    \end{enumerate}
\end{enumerate}
\[ \Xi^{(\Phi)} = \left( \begin{array}{ccc}
  \Enc\Big(k_p, \mu_\kappa\big(\Phi(0, \dots, 0)\big)\Big) & \cdots & \Enc\Big(k_p, \mu_\kappa\big(\Phi(0, \dots, 0)\big)\Big) \\
  \Enc\Big(k_p, \mu_\kappa\big(\Phi(0, \dots, 1)\big)\Big) & \cdots & \Enc\Big(k_p, \mu_\kappa\big(\Phi(0, \dots, 1)\big)\Big) \\
  \vdots & & \vdots \\
  \Enc\Big(k_p, \mu_\kappa\big(\Phi(l_1, \dots, l_K)\big)\Big) & \cdots & \Enc\Big(k_p, \mu_\kappa\big(\Phi(l_1, \dots, l_K)\big)\Big) \\
  \vdots & & \vdots \\
  \Enc\Big(k_p, \mu_\kappa\big(\Phi(m_1, \dots, m_K)\big)\Big) & \cdots & \Enc\Big(k_p, \mu_\kappa\big(\Phi(m_1, \dots, m_K)\big)\Big)
\end{array} \right) \]

\subsection{Component inference}

Since $\chi_1, \dots, \chi_K$ want to keep the component test data secret, inference on component reliability is performed by each manufacturer prior to any encryption.  This corresponds to each $\chi_k$ estimating $\mathbb{P}\left(C^k_t = l_k\right)$ for each $t \in \{ t_1, \dots, t_n \}$.  Given that $\delta$ cannot validate the inference approach taken to estimate $\mathbb{P}\left(C^k_t = l_k\right)$ from the test data, one might argue a non-parametric method should be specified as part of the protocol.  This is the approach used in the example in the next section.

\subsection{The protocol}

The privacy preserving protocol then proceeds as follows:

\begin{enumerate}
  \item $\delta$ sends $k_p$, $\mathbf{t}$ and $\kappa$ to all $\chi_k$.  Each $\chi_k$ is also sent only the $k$-th column of $\Xi^{(l)}$.  Finally, $\Xi^{(\Phi)}$ is sent to $\chi_1$ to start the protocol.
  \item For $k = \{1, \dots, K\}$:
    \begin{enumerate}
      \item For $t = \{1, \dots, T\}$:
        \begin{itemize}
          \item Update \begin{equation} \Xi^{(\Phi)}_{i,t} = \Xi^{(\Phi)}_{i,t}\ \otimes\ \Enc\Big(k_p, \mu_\kappa\big( \mathbb{P}(C^k_t = \Xi^{(l)}_{i,k}) \big)\Big) \quad \forall\,i \label{eq:EncCompUpdate} \end{equation}
        \end{itemize}
      \item After fully updating $\Xi^{(\Phi)}$, if $k<K$ then $\chi_k$ sends $\Xi^{(\Phi)}$ to $\chi_{k+1}$.
    \end{enumerate}
    \item The final participant $\chi_K$ computes the final result, the encrypted column sums of $\Xi^{(\Phi)}$,
      \begin{equation} \xi_j \triangleq \bigoplus_{i=1}^{\prod_{k=1}^K (M_K+1)} \Xi^{(\Phi)}_{i,t} \quad \forall\,j\in\{1,\dots,T\} \label{eq:ColSumXi} \end{equation}
      $\chi_K$ then sends $\mathbf{\xi}=(\xi_1, \dots, \xi_T)$ to $\delta$.
    \item $\delta$ finally computes $10^{-(K+1)\kappa}\,\Dec(k_s, \xi_j) \approx \mathbb{P}(T_S > t_j)$, rendering the survival curve values computed on the required grid of times $\mathbf{t}$.
\end{enumerate}
Note that due to the system lifetime decomposition via the survival signature being a homogeneous polynomial in the survival signature and component survival probabilities, the adjustment required in the final step above is known a priori to be $10^{-(K+1)\kappa}$.

\section{Experimental results}\label{sec:ExperimentalResults}

\begin{figure}
  \centering
    \includegraphics{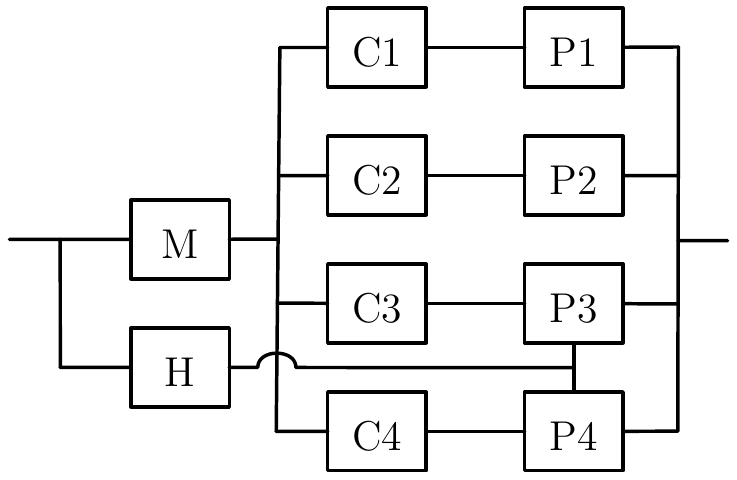}
  \caption{Simple automotive braking system.  The master brake cylinder (M) engages all the four wheel brake cylinders (C1 -- C4).  These in turn each trigger a braking pad assembly (P1 -- P4).  The hand brake (H) goes directly to the rear brake pad assemblies P3 and P4; the vehicle brakes when at least one of the brake pad assemblies is engaged.}
  \label{fig:brake}
\end{figure}

In order to provide an insight into the method for practitioners, a fully encrypted survival analysis was conducted on the simple automotive braking system example from \citep{Walter:2016aa}, depicted in Figure \ref{fig:brake}.  This system comprises four different component types which are taken to be manufactured by different parties, all of whom wish their test data to remain secret.  It is assumed that the system design is also secret and belongs to a different party.  The full code required to run this example is provided in the supplementary material, programmed in the easy to use R language \citep{R-Core-Team:2015aa}.

\subsection{Component inference}

Manufacturer of component $k$ is taken to have $n_k > 0$ observations of lifetime test data on the component type they manufacture, $\mathbf{t}^k = \{ t^k_1, \dots, t^k_{n_k} \}, k \in \{ C, H, M, P \}$.  It is a well studied problem to infer the component lifetime distributions, either non-parametrically or by inferring parameter uncertainty in a family of models such as the Weibull.  \citet{Aslett:2015aa} considered both situations with the survival signature, but in this example only the non-parametric approach is used: both methods would work with the privacy preserving protocol described hereinbefore, but in the present context it would seem unlikely manufacturers would be willing to disclose the parametric family they are fitting and the system designer may therefore wish to guard against inappropriate choices which (s)he cannot validate --- if all manufacturers use the same non-parametric method then the context of all results is well understood.

\subsection{Example setup}

To make the example usage of the technique as realistic as possible, 5 compute servers were launched in 5 different continents around the world using the Amazon Web Services EC2 cloud platform, to mimic a globally distributed supply chain.  This also provides an empirical example of the cost of the method if a designer/manufacturer does not have a sufficiently powerful machine to run the algorithm.  The servers for each party were launched as follows:

\begin{center}
  \begin{tabular}{l@{\hskip 1cm}l@{\hskip 1cm}l}
    \textbf{Role} & \textbf{Physical Server Location} & \textbf{Server Type} \\
    \hline\hline
    System designer & Dublin, Ireland & m4.10xlarge \\
    Manufacturer C & Northern California, USA & m4.10xlarge \\
    Manufacturer H & S\~{a}o Paulo, Brazil & c3.8xlarge\\
    Manufacturer M & Sydney, Australia & r3.4xlarge \\
    Manufacturer P & Tokyo, Japan & i2.8xlarge \\
    \hline
  \end{tabular}
\end{center}

The servers above have the following technical specification:
\begin{center}
  \begin{tabular}{l@{\hskip 0.5cm}l@{\hskip 0.5cm}l@{\hskip 0.5cm}l}
    \textbf{Server Type} & \textbf{Intel Xeon CPU} & \textbf{Memory (GB)} & \textbf{Hourly Cost (US\$)} \\
    \hline\hline
    m4.10xlarge & 40 cores, 2.4 GHz Haswell & 160 & 2.61 -- 2.79 \\
    c3.8xlarge & 32 cores, Ivy Bridge & 60 & 2.60 \\
    r3.4xlarge & 16 cores, Ivy Bridge & 244 & 1.60 \\
    i2.8xlarge & 32 cores, Ivy Bridge & 244 & 8.00 \\
    \hline
  \end{tabular}
\end{center}
`Cores' refers to hyper-threaded cores and hourly costs quoted are on-demand in the region launched.  Much cheaper `spot' prices are commonly available.  See the Amazon Web Services website for further details.

The \citet{Fan:2012aa} homomorphic encryption scheme was used via the HomomorphicEncryption R package \citep{Aslett:2014aa}, with survival signature computed by the system designer using the ReliabilityTheory R package \citep{Aslett:2012aa}.

To ensure reasonable transmission times between continents the cipher texts must be compressed.  Typically, the size of $\Xi^{(\Phi)}$ in this example reached nearly 12GB and although the standard Unix gzip tool provides good compression, it can take up to 20 minutes to compress such a large file.  Therefore pigz \citep{Adler:2015aa} was used for high speed parallel compression which was more than an order of magnitude faster on the multi-core Amazon EC2 servers.

\subsection{The encrypted analysis run}

The system designer set precision $\kappa=5$; set 100 evenly spaced times for evaluation between 0 and 5, $\mathbf{t} = \{ \frac{5i}{99} : i = 0, 1, \dots, 99 \}$; and generated public/secret keypairs providing at least 128-bit security (that is, a brute force attacker would have to perform order $2^{128}$ operations to decrypt).

The computational characteristics of this actual analysis run were as follows:
\begin{center}
  \begin{longtable}{p{5cm}l@{\hskip 0.5cm}rr}
    \textbf{Role} & \textbf{Action} & \multicolumn{2}{l}{\textbf{Timing / Size}} \\
    \endfirsthead
    \textbf{Role} & \textbf{Action} & \multicolumn{2}{l}{\textbf{Timing / Size}} \\
    \hline\hline
    \endhead
    \hline\hline
    \multirow{5}{5cm}{System designer \newline Dublin, Ireland} & Generation of $(k_p, k_s)$ &  & 0.3 secs \\
    & Encryption of $\Xi^{(\Phi)}$ & 1 min & 41.1 secs \\
    & Saving $\Xi^{(\Phi)}$ to disk & 2 min & 41.3 secs \\
    & Compressing $\Xi^{(\Phi)}$ on disk &  & 48.0 secs \\ 
    & Size of $\Xi^{(\Phi)}$ on disk & \multicolumn{2}{c}{5.5GB} \\
    \hline
    \multicolumn{2}{c}{\emph{Transfer $\Xi^{(\Phi)}$ to Manufacturer C}} & 11 min & 37.5 secs \\
    \hline
    \multirow{3}{5cm}{Manufacturer C \newline Northern California, USA} & Decompress \& load $\Xi^{(\Phi)}$ from disk & 10 min & 22.4 secs \\
    & Update $\Xi^{(\Phi)}$, eq \eqref{eq:EncCompUpdate} & 6 min & 18.3 secs \\
    & Saving \& compressing $\Xi^{(\Phi)}$ to disk & 2 min & 9.8 secs \\
    \hline
    \multicolumn{2}{c}{\emph{Transfer $\Xi^{(\Phi)}$ to Manufacturer H}} & 11 min & 24.4 secs \\
    \hline
    \multirow{3}{5cm}{Manufacturer H \newline S\~{a}o Paulo, Brazil} & Decompress \& load $\Xi^{(\Phi)}$ from disk & 10 min & 13.2 secs \\
    & Update $\Xi^{(\Phi)}$, eq \eqref{eq:EncCompUpdate} & 7 min & 23.1 secs \\
    \nopagebreak[3]& Saving \& compressing $\Xi^{(\Phi)}$ to disk & 4 min & 45.2 secs \\
    \hline
    \multicolumn{2}{c}{\emph{Transfer $\Xi^{(\Phi)}$ to Manufacturer H}} & 20 min & 16.5 secs \\
    \hline
    \multirow{3}{5cm}{Manufacturer M \newline Sydney, Australia} & Decompress \& load $\Xi^{(\Phi)}$ from disk & 9 min & 41.0 secs \\
    & Update $\Xi^{(\Phi)}$, eq \eqref{eq:EncCompUpdate} & 11 min & 28.2 secs \\
    & Saving \& compressing $\Xi^{(\Phi)}$ to disk & 2 min & 54.2 secs \\
    \hline
    \multicolumn{2}{c}{\emph{Transfer $\Xi^{(\Phi)}$ to Manufacturer H}} & 6 min & 40.7 secs \\
    \hline
    \multirow{5}{5cm}{Manufacturer P \newline Tokyo, Japan} & Decompress \& load $\Xi^{(\Phi)}$ from disk & 9 min & 57.1 secs \\
    & Update $\Xi^{(\Phi)}$, eq \eqref{eq:EncCompUpdate} & 7 min & 13.5 secs \\
    & Compute $\xi$, eq \eqref{eq:ColSumXi} & & 6.1 secs \\
    & Saving \& compressing $\xi$ to disk &  & 2.5 secs \\
    & Size of $\xi$ on disk & \multicolumn{2}{c}{58.4MB} \\
    \hline
    \multicolumn{2}{c}{\emph{Transfer $\xi$ to System Designer}} & & 39.5 secs \\
    \hline
    \multirow{2}{5cm}{System designer \newline Dublin, Ireland} & Decompress \& load $\xi$ from disk &  & 5.9 secs \\
    & Decryption of $\xi$ &  & 8.6 secs \\
    \hline\hline
    \textbf{Total:} & \multicolumn{1}{r}{2 hr} & 18 min & 38.4 secs
  \end{longtable}
\end{center}

\begin{figure}
  \centering
    \includegraphics{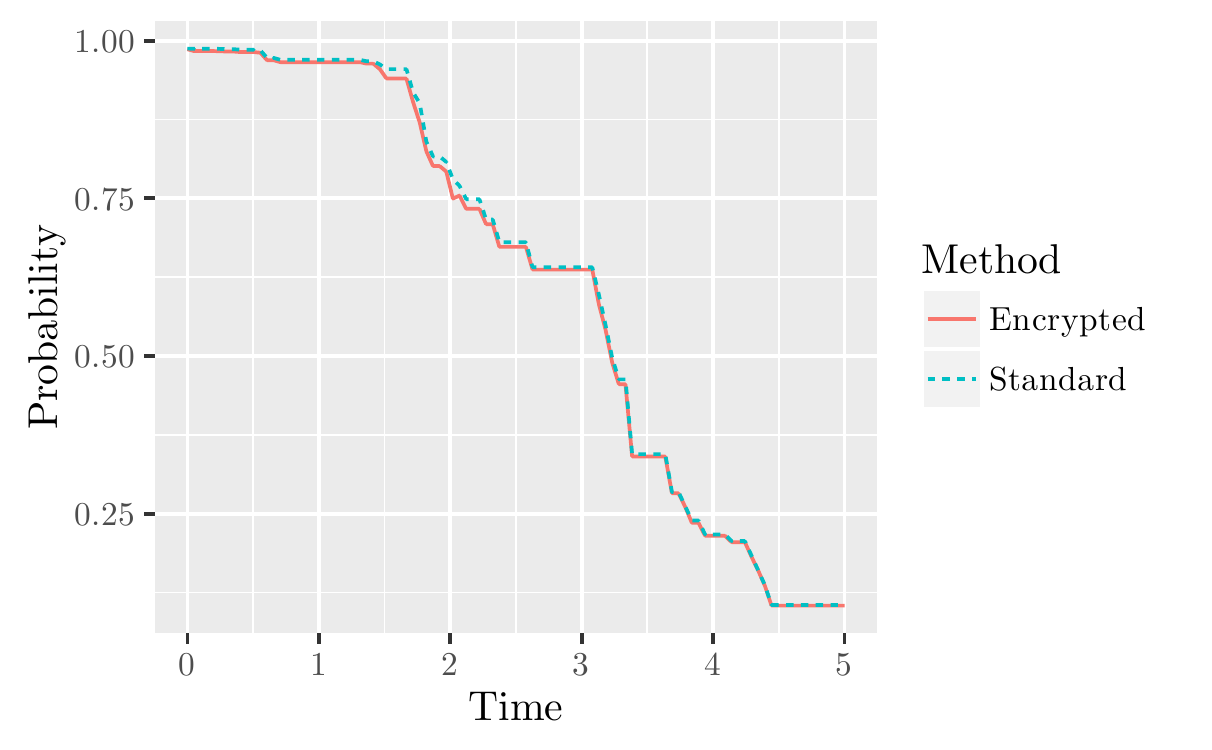}
  \caption{Comparison of the system survival curves as estimated using the software function in \citet{Aslett:2015aa} and as decrypted after the encrypted estimation on the global network of servers.}
  \label{fig:survcurve}
\end{figure}

Clearly the most variable timing element of this analysis stems from the data transfer time between manufacturers, which ranged from under 7 minutes to over 20 minutes.  A total runtime of under 3 hours means that this protocol is entirely feasible to run even if manufacturer or designer need to outsource the compute power to accomplish it.  In particular, none of the machines needs to be on for more than 1 hour and at the time of writing the most expensive of the instances used costs US\$\,7.50 per hour, though using so-called spot-instances there is usually a suitable instance type available for less than US\$\,1 per hour.

Upon decryption, the survival curve which is yielded can be seen in Figure \ref{fig:survcurve}.  Compared to the baseline result which is computed unencrypted, there is a very small discrepancy resulting from the rounding associated with the integer encoding $\mu_\kappa: \mathbb{R} \to \mathbb{Z}$.  An empirical estimate of the total variation distance between the true and encrypted survival probability measures is 0.029, representing a small discrepancy.

\section{Conclusions}\label{sec:Conclusions}

This work has developed a privacy preserving protocol for conducting a full system reliability analysis.  The use of homomorphic encryption methods provides cryptographically strong guarantees on the security of the raw system design and component test data.

The method has been demonstrated to be practical despite the tremendous computational burden which traditionally accompanies encrypted statistical analysis and is a method which can be implemented by businesses even where they lack in-house compute servers by availing of inexpensive cloud computing options (where privacy is still preserved since all material is encrypted).

\section*{Supplementary materials}

There are extensive supplementary materials provided in a compressed zip file.  The contents are as follows:

\begin{description}
  \item[\texttt{designer.R}:] An extensively commented script file containing the code which a system designer would run in order derive the survival signature and homomorphically encrypt $\Xi^{(\Phi)}$ as described in Section \ref{sec:PrivacyPreservingProtocol}, using the automotive braking system of Section \ref{sec:ExperimentalResults}. (R script file)
  \item[\texttt{manufacturer_\{C,H,M,P\}.R}:] Four script files, one per manufacturer, containing the code in order to execute the steps of the protocol described in \ref{sec:PrivacyPreservingProtocol} for the braking system example of Section \ref{sec:ExperimentalResults}. (R script file)
  \item[\texttt{utility.R}:] defines utility functions used in the above scripts to compute $\mu_\kappa(\cdot)$, $\mu_\kappa^{-1}(\cdot)$ and to compute the non-parametric component inference estimate.  (R script file)
  \item[\texttt{AmazonEC2-xi_m.RData}:] the decrypted results of the full experiment run in Section \ref{sec:ExperimentalResults}.  (R image file)
  \item[\texttt{compare_to_truth.R}:] a script which plots the experimental results which were run encrypted across a global network to the result of performing the experiment unencrypted in one session.  (R script file)
\end{description}

\section*{Acknowledgements}

The author is supported by the i-like programme grant (EPSRC grant reference number EP/K014463/1 http://www.i-like.org.uk).

\bibliographystyle{abbrvnat}
\bibliography{References}

\end{document}